\documentclass[prb, amssymb, showpacs, twocolumn]{revtex4}
\usepackage{graphicx}
\usepackage{bm}
\begin{document}
\title{Variational approach to vortex penetration and vortex interaction}
\author{Alexander~D.~Hern\'{a}ndez}
\email{alexande@cab.cnea.gov.ar}
\affiliation{Centro At\'{o}mico Bariloche and Instituto Balseiro, (8400) Bariloche, Argentina}
\author{Arturo~L\'{o}pez}
\affiliation{Centro At\'{o}mico Bariloche and Instituto Balseiro, (8400) Bariloche, Argentina}
\date{\today}

\begin{abstract}
A variational calculation for vortex penetration is presented.
Variational trial functions for the Meissner state are combined with variational
functions for a vortex near the surface. The latter is based on
Clem's trial solutions for a vortex in bulk, which were adapted to include
surface effects through consideration of an image vortex.
Three variational parameters are
considered, corresponding to the effective coherence length of the
vortex, the effective penetration length for the Meissner currents,
and the value of the order parameter at the surface. The results
show that the last two variational parameters are independent of
vortex position. Explicit calculations are presented for
several $\kappa$ values. The energy barrier for vortex
penetration is shown to be in good agreement with full numerical calculations of the
Ginzburg--Landau equations. We consider the variation of the magnetic flux carried
by a vortex  as it gets inside the superconductor and agreement with known experimental and theoretical results is obtained.
The model was extended to calculate the force between two vortices, and the results show that the force goes to zero as the pair comes close to the surface. This result can be of interest for the study of the melting of the vortex lattice and for vortices confined in mesoscopic superconductors. The variational approach can be very helpful for intermediate $\kappa$ values when numerical calculations become computationally demanding because it provides manageable expressions for all physically relevant quantities.

\end{abstract}

\pacs{74.20.De, 74.25.Ha, 74.25.Op} \maketitle

\section{Introduction}

The behavior of a vortex near the surface of a superconductor has
been the subject of several recent papers. \cite{zeldov, wei, geim1,
singha, mingo1, golib} Semi analytical results have been known since the
Bean--Livingston\cite{bean} model was formulated and simple
calculations were made from the London model by de Gennes.
\cite{degennes} Among other properties, these calculations give the
characteristics of the surface barrier for vortex penetration. The
geometrical  surface barrier in superconducting thin films has been
considered for high-$T_c$ superconductors in Refs.
\onlinecite{zeldov} and \onlinecite{wei}. While the Bean Livingston barrier is of
energetic origin, the geometrical barrier is strongly dependent on
sample shape. The surface barriers are also very important in
mesoscopic superconductors.\cite{geim1, singha, mingo1, golib}
Interesting results pertaining to the ac magnetic properties of mesoscopic superconductors 
have been obtained from numerical calculations based on the 
finite-difference method in Refs. \onlinecite{mingo1} and \onlinecite{mingo1a}.

On the other hand, variational calculations are known to provide manageable and
accurate results in many different physical problems. In other
applications of the Ginzburg--Landau (GL) equations, variational
calculations are known to give good agreement with exact results. In
some cases, variational calculations have preceded exact or numerical
calculations. Such is the case for surface superconductivity,\cite{degennes,tinkham}
the mixed state in type II superconductors, \cite{abrikosov} and, more recently, superconducting
micronetworks.\cite{castro}

In this paper, we present a variational approach to the solution of
the GL equations for a vortex near the surface of a superconductor,
starting from Clem's variational ansatz\cite{clem} for a vortex in
bulk. In this form, we are able to compare the variational results to the 
full numerical calculations. Vortices appear in the presence of an
externally applied field, which also induces Meissner currents. For
this reason, it is necessary to variationally model both aspects of the behavior of
a superconductor. The Clem ansatz has also been used recently in
the context of mesoscopic superconductors in Ref. \onlinecite{cabral}.

The paper is organized as follows: In Sec. II.A, we present  a
variational description of the Meissner state, and the variational solution is compared to the  full numerical results of the GL equations in Sec. II B. In Sec. III, Clem's
ansatz for a single vortex in a bulk material is adapted to describe
a vortex near the surface of a superconductor and is combined with
the description of the Meissner state. In Secs. III A and III B, we present the
results of the variational calculations including these three parameters: the
penetration length for the Meissner currents, the order parameter at
the sample surface, and the coherence length for the vortex size. It
turns out that the first two parameters are independent of the
vortex position. The results of the variational calculation are
compared to the full numerical results, particularly for the
energy barrier for $\kappa=2$ and $\kappa=3$. Quite reasonable
agreement between both methods is obtained for both $\kappa$ values. In Sec. IV, the model is extended to calculate the force between
two vortices as a function of their distance to the surface. We show that the interaction force
goes to zero as the pair approaches the sample surface. Finally, in Sec. V, we give our conclusions.

\section{Description of the Meissner state}

In this section, we propose a variational model to describe the
Meissner state of a semi-infinite sample. In Sec. II A we obtain an approximate solution of the GL equations valid at low magnetic fields when depletion of the order parameter at the sample surface is small. By using this approximate solution, we propose a variational model to describe the Meissner state at higher values of the field. In Sec. II B, the variational solution is compared to the full numerical results of the GL equations.

\subsection{Variational model for the Meissner state}

By writing the order parameter as $\psi =fe^{i\varphi }$, we obtain the following expression in normalized units for the difference between the free energy of the normal and superconducting states (${\cal G}_{s}-{\cal G}_{n}$):

\begin{eqnarray}
{\cal G}_{s}-{\cal G}_{n}&=&\int\Big[-f^{2}+\frac{f^{4}}{2}+\frac{1}{
\kappa ^{2}}\left[ (\nabla f)^{2}+(\nabla \varphi -\kappa {\bf A})^{2}f^{2}\right] \nonumber \\  
&+&({\bf B}-{\bf H}_{a})^{2}\Big] d^{3}r. \label{1}
\end{eqnarray}

Lengths ${\bf r}$ are scaled in units of the zero temperature penetration length $
\lambda (0)$, the externally applied magnetic field ${\bf H}_a$ and ${\bf B}$ also in units of $\sqrt{2}H_{c}(0)$, the vector potential ${\bf A}$ in units of $\sqrt{2}\lambda (0)H_{c}(0)$, the order parameter $\psi $ in units of $\psi _{\infty }$, the current ${\bf J}$ in units of $\Psi_{\infty }^{2}e\hbar /m\xi $, and velocities in units of $\hbar
/2m\xi(0)$.

The GL equations become

\begin{eqnarray}
\frac{1}{\kappa ^{2}}\nabla ^{2}f &=&f\left( f^{2}+\frac{1}{\kappa ^{2}}%
\left( \nabla \varphi -\kappa {\bf A}\right) ^{2}-1\right),\label{2} \\
&&  \nonumber \\
\nabla \times \nabla \times {\bf A} &=&f^{2}\left( \frac{\nabla \varphi }{%
\kappa }-{\bf A}\right).  \label{3}
\end{eqnarray}

We assume a semi-infinite medium subjected to a magnetic field
parallel to the superconductor-vacuum interface. We choose the
$\hat{x}$ axis perpendicular to this interface and take the
$\hat{z}$ direction parallel to the applied field ${\bf
B}=B_{z}(x){\hat{z}}$.

Equations (\ref{2}) and (\ref{3}) must be complemented with the 
appropriate boundary conditions at the sample surface, which when 
separated into its real and imaginary parts, imply $(\nabla f)_\perp|_{s}=0$ and
$ ({\bf u})_\perp|_{s}=0$, where the first relation indicates that the slope of the order parameter perpendicular to the surface must be zero at the surface, whereas the second implies that the velocity of the superconducting electrons (${\bf u}=(\nabla \varphi -\kappa {\bf A})$) has no component perpendicular to the surface. For a semi-infinite sample with no demagnetizing effects, the condition ${\bf B}|_{s}={\bf H}_{a}$, where ${\bf H}_{a}$ is the externally applied field, also applies at the surface.

 In this configuration, the order parameter depends only on $x$, $f=f(x)$. Moreover, $ \nabla \times {\bf B}$ has only a non-vanishing component, $\left( \nabla \times {\bf B}\right)
_{y}=\partial _{z}B_{x}-\partial _{x}B_{z}=-\partial
_{x}B_{z}(x){\hat{y}}$. From Eq. (\ref{3}), we then have ${\bf
A}=A_{y}(x){\hat{y}}$.

In the London gauge, the order parameter is real and we can eliminate the phase $\varphi$ in the  GL equations. In this geometry and specific gauge, Eqs. (\ref{2}) and (\ref{3}) become

\begin{eqnarray}
\frac{1}{\kappa ^{2}}\frac{d^{2}f}{dx^{2}} &=&f\left( f^{2}-1\right) +\left(
A_{y}\right) ^{2}f,  \label{4} \\
&&  \nonumber \\
\frac{d^{2}A_{y}}{dx^{2}} &=&f^{2}A_{y},  \label{5}
\end{eqnarray}
with the following boundary conditions: $(df/dx)|_{x=0}=0$ and $%
{\bf B}|_{x=0}=(dA_{y}/dx)|_{x=0}=H_{a}$.

An approximate solution of these equations at low fields can be found by assuming $f(x)=f_{\infty }-\eta (x)$, with $|\eta (x)|\ll f_{\infty }$. In the present normalization,
$f_{\infty }=1$; thus, we can write the solution to Eq. (\ref{5}) when
$\eta (x) \rightarrow 0$ as

\begin{equation}
A_{y}\approx -H_{a}e^{-x}.  \label{7}
\end{equation}

In the following, we assume that at low fields for $f(x)\lesssim 1$, the vector potential can
be conveniently approximated by a variational expression of the following form:
\begin{equation}
A_{y}=-\lambda _{M}H_{a}e^{-x/\lambda _{M}},\label{8}
\end{equation}
where $\lambda _{M}$ is a variational parameter. As will be the case
for the other two variational parameters to be introduced later,
$\lambda _{M}$ is a field and temperature dependent parameter.

By using Eq. (\ref{8}) and $f(x)=1-\eta(x)$, in a first approximation, Eq. (\ref{4}) becomes 

\begin{equation}
\frac{1}{\kappa ^{2}}\frac{d^{2}\eta }{dx^{2}}=2\eta
-H_{a}^{2}\lambda _{M}e^{-2x/\lambda _{M}}.  \label{9}
\end{equation}

By solving Eq. (\ref{9}) with the boundary conditions $(d\eta
/dx)_{x=0}=0$ and $\eta |_{x\rightarrow \infty }=0$, we obtain the following expression for
the depletion of the order parameter:

\begin{equation}
\eta (x)=\frac{\eta _{0}}{(\kappa \lambda _{M}-\sqrt{2})}\left(
\kappa \lambda _{M}e^{-2x/\lambda _{M}}-\sqrt{2}e^{-\sqrt{2}\kappa
x}\right), \label{eta1}
\end{equation}
where $\eta _{0}$ is the value of $\eta (x)$ at the sample surface,
\begin{equation}
\eta (0)=\eta _{0}=\frac{H_{a}^{2}\kappa\lambda _{M} }{2(\kappa \lambda _{M}+%
\sqrt{2})}\label{eta2}.
\end{equation}
This relation is complemented by the expression for the magnetic
field $B_{z}(x)$,

\begin{equation}
B_{z}(x)=H_{a}e^{-x/\lambda _{M}},  \label{10}
\end{equation}
which follows at once from Eq. (\ref{7}). Both $\eta (x)$ and $B_{z}(x)$ depend on the variational parameter $\lambda _{M}$, which can be obtained by minimizing the Gibbs free energy
[Eq. (\ref{1})].

In Eq. (\ref{eta1}), $\eta _{0}$ is related to $\lambda_{M}$ through Eq. (\ref{eta2}).
The approximation is more accurate the lower the magnetic field is, i.e., when $\eta _{0}\ll 1$.
In the above equations, we have only one variational parameter, which is $\lambda_{M}$.
An alternative possibility is to consider $\eta _{0}$ as a second variational parameter
to obtain a more accurate description of the Meissner state up to magnetic fields
close to the vortex penetration field $H_{p}$. We have followed this second procedure in this paper.

To determine the variational parameters, we must find the extremum of 
$\frac{\Delta {\cal G}}{L_{z}L_{y}}$ given by Eq. (\ref{1}), which can be
written as

\begin{eqnarray}
\frac{\Delta {\cal G}}{L_{z}L_{y}}=&&\int_{0}^{\infty }dx\left( -f^{2}+\frac{f^{4}}{2}+\frac{1}{\kappa ^{2}}%
\left( \left( \nabla f\right) ^{2}+{\bf u}^{2}f^{2}\right) \right) \nonumber \\ 
&+&\int_{0}^{\infty }dx[{\bf B}(x)-{\bf H}_{a}]^{2}, \label{13}
\end{eqnarray}

where ${\bf u}$ is the velocity of the superconducting electrons, ${\bf u}%
=(\nabla \varphi -\kappa {\bf A})$. In the London gauge, ${\bf u}$ is
proportional to ${\bf A}$, ${\bf u}=-\kappa {\bf A}$; therefore, we have
\begin{eqnarray}
u_x&=&0, \nonumber \\
u_y&=&-\kappa\lambda _{M}H_{a}e^{-x/\lambda _{M}}.\label{13a}
\end{eqnarray}
The free energy [Eq.  (\ref{13})] can be evaluated by using Eqs. (\ref{eta1}), (\ref{10}), and (\ref{13a}) for $f(x)=1-\eta(x)$, $B(x)$, and $u$, respectively. The minimization of the free energy allows us to obtain $\lambda _{M}$ and $\eta _{0}$, which completes the variational description of the Meissner state.

\subsection{Full numerical time-dependent Ginzburg--Landau solution}

We compare the variational solutions to the results obtained from
full numerical solutions of the time-dependent GL equations,\cite{groop,kato,bolech}

\begin{eqnarray}
\frac{\partial \Psi}{\partial t} = \frac{1}{\kappa^2}(\nabla -
i\kappa A)^2 \Psi +(1-|\Psi |^2)\Psi,    \label{TDGL1} \\
\frac{\partial A}{\partial t} = \frac{1}{\sigma'} \left(\frac{\mbox{Im}[\Psi^* (\nabla -
i \kappa A)\Psi]}{\kappa} -\nabla \times \nabla \times A \right). \label{TDGL2}
\end{eqnarray}

Time is in the units of a characteristic normalization time $\tau=\xi^2/D$, where $D$ is the electronic diffusion constant. $\sigma'$ is the normalized conductivity,  $\sigma'=(4\pi\lambda^2/c^2\tau)\sigma$. The other quantities were normalized as in the previous section.

 To solve equations (\ref{TDGL1}) and (\ref{TDGL2}), we have used the standard finite-difference discretization scheme. \cite{groop} The order parameter and vector potential are defined at the nodes of a rectangular mesh (${\bf r}=(I,J)$).
In our simulations, we have assumed a sample that is semi-infinite in
the $x$ direction and infinite in the $y$, $z$ directions, and we have assumed that 
the magnetic field is applied along $z$. The problem is then reduced to
two dimensions because we can neglect all derivatives along $z$. The
symmetry of the problem implies that for all mesh points,
${\bf A}_{I,J}=(A_{xI,J},A_{yI,J},0)$ and
 ${\bf B}_{I,J}=(0,0,B_{zI,J})$, where
 $B_{_zI,J}=(\nabla \times {\bf A})_z=(\partial_xA_{_yI,J}-\partial_yA_{_xI,J})$.
The link variables $U_{_\mu I,J}=\exp(-\imath \kappa h_{\mu} A_{\mu I,J})
\;\; (\mu=x,y)$ are introduced in order to preserve gauge invariance in the discretization.

In this geometry, the discretized forms of equations (\ref{TDGL1})
and (\ref{TDGL2}) are 
\begin{eqnarray} \nonumber
\frac{\partial \Psi}{\partial
t}&=&\frac{U^*_{_xI-1,J}\Psi_{I-1,J}-2\Psi_{I,J}+U_{_xI,J}\Psi_{I+1,J}}{(\kappa \Delta x)^2} \nonumber \\
  &+&
 \frac{U^*_{_yI,J-1}\Psi_{I,J-1}-2\Psi_{I,J}+U_{_yI,J}\Psi_{I,J+1}}{(\kappa \Delta y)^2}  \nonumber\\
&+&(1-|\Psi_{I,J}|^2)\Psi_{I,J}, \\
\frac{\partial A_{_xI,J}}{\partial t}&=&\frac{1}{\sigma'}
\left(\frac{\mbox{Im} [U_{_xI,J}\Psi^*_{I,J} \Psi_{I+1,J}]}{\kappa \Delta x} -
\frac{B_{_zI,J}-B_{_zI,J-1}}{\Delta y}\right),  \\
\frac{\partial A_{_yI,J}}{\partial t}&=&\frac{1}{\sigma'}\left(\frac{\mbox{Im}
[U_{_yI,J}\Psi^*_{I,J} \Psi_{I,J+1}]}{\kappa \Delta y}
+\frac{B_{_zI,J}-B_{_zI-1,J}}{\Delta x}\right),
\end{eqnarray}

where $\Delta x$ and $\Delta y$ are the mesh widths. $\sigma'$ was chosen as equal to unity, as in Ref. \onlinecite{groop}.

The dynamical equations must be complemented with the appropriate
boundary conditions for both the order parameter and the vector
potential. We have imposed periodic boundary conditions in the
$\hat{y}$ direction, i.e.,
\begin{eqnarray}
\Psi(x,y)&=&\Psi(x,y+L_y), \nonumber \\
A_x (x,y)&=&A_x (x,y+L_y), \nonumber \\
A_y (x,y)&=&A_y (x,y+L_y), \nonumber
\end{eqnarray}
and semiperiodic boundary conditions in the $\hat{x}$ direction, 
where one side of the superconductor is in contact with the vacuum
at $x=0$, implying

\begin{eqnarray}
((\nabla -i{\bf A}) \Psi)_\perp |_{x=0}&=&0, \nonumber \\
B|_{x=0}&=&H_a. \nonumber
\end{eqnarray}

At $x=L$, we impose the conditions that are obtained at $x=\infty$,

\begin{eqnarray}
|\Psi|^2 |_{x=L_x}&=&1, \nonumber \\
B|_{x=L_x}&=&0. \nonumber
\end{eqnarray}

By choosing a value much larger than $\lambda$, $L_x=24\lambda$ for $L_x$,
we have obtained accurate results for a sample semi-infinite in
$\hat{x}$.

\subsection{Comparison between the variational solution and the full Ginzburg--Landau numerical results}

\begin{figure}[htb]
\begin{center}
\includegraphics[width=\linewidth]{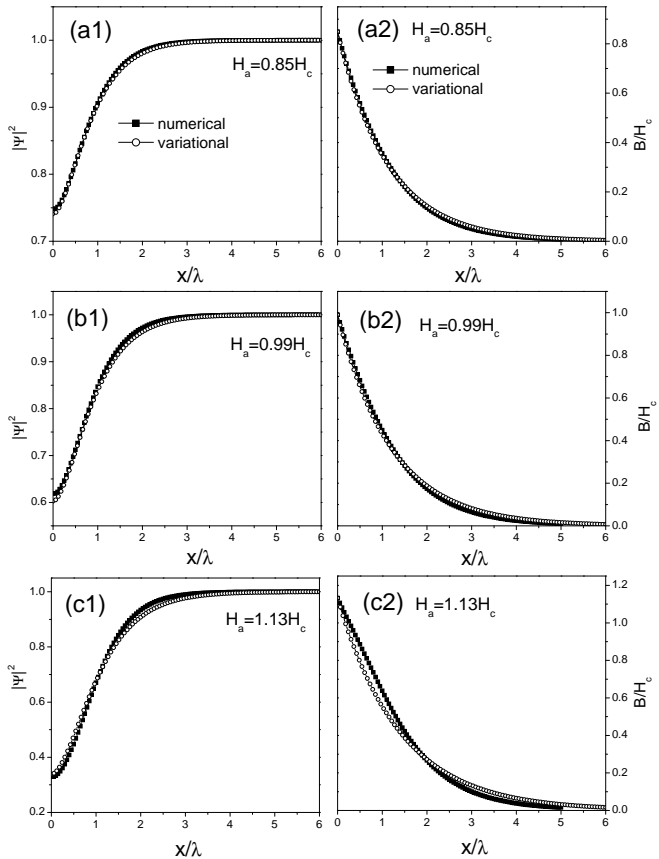}
\caption{ Shown is a comparison of the order parameter and
the magnetic field, in the variational approximation and in the
numerical calculation, for $\kappa =2$ and for different
values of the applied field. We see that even in (c1) and (c2), when $H$ 
is near the field of first penetration, the variational description
is quite accurate.} \label{meissner}
\end{center}
\end{figure}

Figure \ref{meissner} shows a comparison between the variational and full
numerical results for the order parameter and for the magnetic field
in the Meissner state. Both quantities are calculated along a
direction perpendicular to the sample surface and for $\kappa =2$.
The size of the numerical sample is described by $L_x=24\lambda$  and
$L_y=16\lambda$. It is seen that the variational description is
quite accurate, even when $H_a$ is near $H_p$, the field of first vortex
penetration [see Figs. \ref{meissner}(c1) and \ref{meissner}(c2)]. The numerical simulations obtain $H_{p}=1.13H_{c}$; this value coincides with the results of Ref.
\onlinecite{matricon}, which is also a one-dimensional calculation. In a
two-dimensional sample, this result would be obtained for a perfect
surface that induces a uniform depletion of the order parameter
along the surface (see Ref. \onlinecite{carty}  for a complete
discussion). When a defect\cite{vodolazov} or thermal fluctuations\cite{bolech} induce the nucleation of a vortex, the value of $H_{p}$ diminishes.

\section{Clem's variational solution near a surface}

Originally developed for electrostatics and fluid dynamics, the image method
was envisaged to automatically satisfy the boundary conditions in a given
problem. It has found applications in several fields of physics described by
linear equations, wherein the superposition principle is valid. In such cases,
the method provides the exact solution by adding the fields produced by the
real charge and by the image charges.

Since the GL equations are non-linear, care must be exercised in applying
the method. To perform a variational calculation in the present case, we must physically construct  acceptable trial functions for the order parameter and currents.

Clem's \cite{clem} variational calculation allows us to determine the order
parameter, field, and current for a vortex in an infinite superconductor. In
order to use this solution for a vortex close to the superconductor-vacuum
interface, we must first consider a vortex placed at a generic point $\left(
x_{0},y_{0}\right) $ in a bulk superconductor. We introduce the following auxiliary
variables:
\begin{equation}
\begin{array}{ll}
\rho \left( x_{0},y_{0};x,y\right) = & \sqrt{\left( x-x_{0}\right)
^{2}+\left( y-y_{0}\right) ^{2}}, \\
&  \\
R\left( x_{0},y_{0};x,y\right) = & \sqrt{\left( x-x_{0}\right) ^{2}+\left(
y-y_{0}\right) ^{2}+\zeta _{v}^{2}}.
\end{array}
\label{rhoyr2}
\end{equation}

Here, $\zeta _{v}$ is a variational parameter of the same order of magnitude of the coherence
length. Clem's variational ansatz for the order parameter takes the form
\begin{equation}
f_{vor}\left( x,y\right) =\frac{\rho \left( x_{0},y_{0};x,y\right) }{R\left(
x_{0},y_{0};x,y\right) }.
\end{equation}
This allows the exact solution of the second GL equation, giving, respectively, for the
field and current

\begin{eqnarray}
B_{z}&=&  \frac{1}{\kappa \zeta _{v}}\frac{K_{0}(R\left(
x_{0},y_{0};x,y\right) )}{K_{1}(\zeta _{v})}, \nonumber \\
j_{\varphi }&=& \frac{1}{\kappa \zeta _{v}}\frac{\rho \left(
x_{0},y_{0};x,y\right) }{R\left( x_{0},y_{0};x,y\right) }\frac{K_{1}(R\left(
x_{0},y_{0};x,y\right) )}{K_{1}(\zeta _{v})},
\label{clem1}
\end{eqnarray}
where $K_{0}(x)$ and $K_{1}(x)$ are modified Bessel functions.

As we saw above, the boundary conditions at the sample surface require the order parameter
to have a zero slope there. A second requirement is that no current should flow
across the sample surface,
\begin{equation}
\begin{array}{rl}
\left. \frac{df\left( x,y\right) }{dx}\right| _{x=0}= & 0, \\
&  \\
\left. J_{x}\left( x,y\right) \right| _{x=0}= & 0. \\
&  \\
\end{array}
\label{boundary}
\end{equation}
Both conditions can be satisfied by considering the combined effect of the
vortex located at the point ($x_{0},y_{0}$) plus an image vortex, which is located at the point $(-x_{0},y_{0})$. The
currents in the image vortex must rotate in the opposite sense to the ones
in the real vortex. The order parameter for the image vortex would be

\begin{equation}
f_{im}\left( x,y;x_{0},y_{0}\right) =\frac{\rho \left(
x,y;-x_{0},y_{0}\right) }{\sqrt{\rho \left( x,y;-x_{0},y_{0}\right)
^{2}+\zeta _{v}^{2}}}.  \label{orderparamtotal}
\end{equation}

To construct the variational order parameter for the vortex-image
vortex pair, we can simply take the product
\begin{equation}
F( x,y;x_{0},y_{0}) =f_{vor}( x,y;x_{0},y_{0})
\times f_{im}( x,y;-x_{0},y_{0}). \label{orderpair}
\end{equation}

This assumption guarantees the vanishing of the order parameter at
each vortex core and also of the normal slope at the surface,
\[
\left. \frac{dF\left( x,y\right) }{dx}\right| _{x=0}=0.
\]
When the vortex is placed far enough from the surface, Eq.
(\ref{orderpair}) tends to the correct limits.

The variational solution for the current requires some care. A
 velocity field obtained by adding the current
fields for vortex and image vortex would satisfy the boundary
condition at the superconductor-vacuum interface but would violate
the requirement that the current at each vortex core vanishes. An
alternative is to construct first a compound velocity field by
adding the velocity fields of each vortex. The resulting field would
also satisfy the boundary condition, with the advantage that the
singularities at each vortex core would be maintained, very much as is
the case for the charge-image charge pair in electrostatics. The
total current field must then be calculated from this velocity
field. To obtain the velocity distribution for the vortex-image
vortex system, we must consider the sum of the velocities for each of
these elements as follows:

\begin{eqnarray}
U_{x}\left( x,y\right) = & u_{x}\left( x,y;x_{0},y_{0}\right)
-u_{x}\left( x,y;-x_{0},y_{0}\right)\label{Ux},\\
& \nonumber \\
U_{y}\left( x,y\right) = & u_{y}\left( x,y;x_{0},y_{0}\right)
+u_{y}\left( x,y;-x_{0},y_{0}\right)\label{Uy},
\end{eqnarray}

with
\begin{eqnarray}
u_{x}\left( x,y;x_{0},y_{0}\right) =&-&\frac{1}{\kappa \zeta _{v}}\frac{
K_{1}\left( R\left( x,y;x_{0},y_{0}\right) \right) }{K_{1}(\zeta
_{v})} \nonumber \\
&\times& \left( \frac{\left( y-y_{0}\right) R\left( x,y;x_{0},y_{0}\right) }{
\rho ^{2}\left( x,y;x_{0},y_{0}\right) }\right), \nonumber \\
u_{y}\left( x,y;x_{0},y_{0}\right) =&& \frac{1}{\kappa \zeta _{v}}\frac{
K_{1}\left( R\left( x,y;x_{0},y_{0}\right) \right) }{K_{1}(\zeta
_{v})} \nonumber \\
&\times& \left( \frac{\left( x-x_{0}\right) R\left( x,y;x_{0},y_{0}\right) }{
\rho ^{2}\left( x,y;x_{0},y_{0}\right) }\right). \nonumber
\end{eqnarray}

In these expressions, $u_{x}\left( x,y;x_{0},y_{0}\right)$ and
$u_{y}\left( x,y;-x_{0},y_{0}\right)$ are the velocity field
components of a vortex centered at $\left(x_{0},y_{0}\right)$,
whereas $u_{x}\left( x,y;-x_{0},y_{0}\right)$ and
$u_{y}\left(x,y;-x_{0},y_{0}\right)$ are the velocity components of
the image vortex centered at $\left( -x_{0},y_{0}\right)$.

This combination keeps the essential property of having the correct
divergence at each vortex core, and it satisfies the boundary condition

\[
U_{x}\left( 0,y\right) =0.
\]

In order to obtain the variational current field, we must combine the
velocity field of Eqs. (\ref{Ux}) and (\ref{Uy}) with the order parameter for the vortex-image-vortex pair given by Eq. (\ref{orderpair}).
The total current would thus be

\[
{\bf J}\left( x,y\right) =\left| F\left( x,y\right) \right| ^{2}{\bf U}%
\left( x,y\right).
\]
It can be seen that this expression satisfies the correct boundary condition as follows:

\[
J_{x}\left( 0,y\right) =0.
\]

The total magnetic field of the vortex-antivortex pair would be
\[
B_{z}\left( x,y\right) =B_{z}\left( x,y;x_{0},y_{0}\right)
-B_{z}\left( x,y;-x_{0},y_{0}\right)
\]
or
\begin{equation}
B_{z}\left(x,y\right) =\frac{1}{\kappa \zeta _{v}}\left( \frac{%
K_{0}(R\left( x,y;x_{0},y_{0}\right) )}{K_{1}(\zeta _{v})}-\frac{%
K_{0}(R\left( x,y;-x_{0},y_{0}\right) )}{K_{1}(\zeta _{v})}%
\right). \label{BVortAntivor}
\end{equation}

\begin{figure}
\begin{center}
\includegraphics[width=\linewidth]{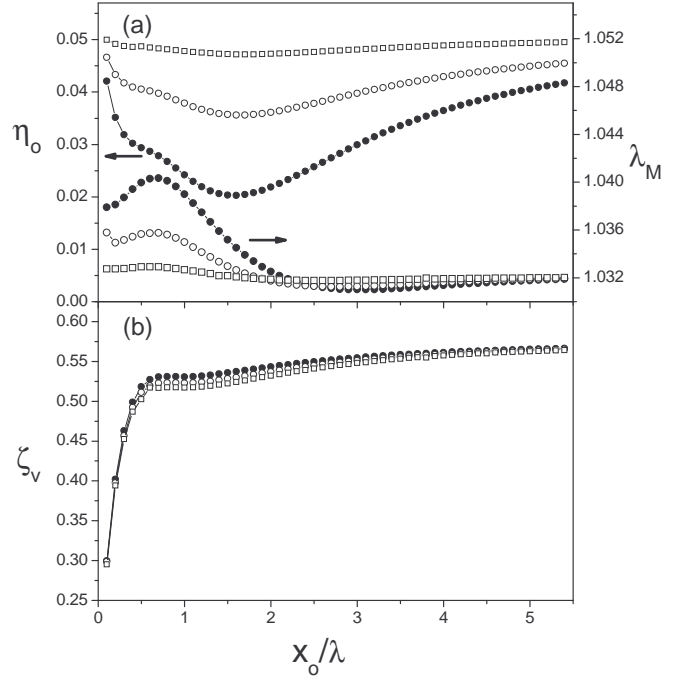}
\end{center}
\caption{{\small Variational parameters as a function of the vortex
position. In (a), parameters $\eta _{0}$ and $\lambda _{M}$, which are related to the Meissner state, are shown. Same as in (b) for the parameter $\zeta _{v}$, which determines the vortex size. System sizes are: ($\bullet$) $20\lambda \times 11 \lambda$, ($\circ$)
$40\lambda \times 11 \lambda$, and ($\square$) $160\lambda \times 11 \lambda$.}}
\label{parameters}
\end{figure}

To evaluate the free energy given by Eq. (\ref{13}), we have to combine the vortex-antivortex expressions with the contributions due to the Meissner currents. For the magnetic field and currents, we assume a superposition principle and simply add the contributions due to each source. Thus, to obtain the total magnetic field, we have to add Eqs. (\ref{10}) and (\ref{BVortAntivor}). The velocity of the superconducting electrons $u$ is obtained by adding the vector component given by Eq. (\ref{13a}) and Eqs. (\ref{Ux}) and (\ref{Uy}). For the order parameter, instead, we have to multiply both contributions, given by Eq. (\ref{orderpair}) and by the contribution of the Meissner state, $f(x)=1-\eta(x)$, with $\eta(x)$ given by
Eq. (\ref{eta1}). The minimization of the free energy allows us to obtain $\zeta _{v}$, $\lambda _{M}$, and $\eta_{0}$, which completes the variational description of a vortex near a surface.

\subsection{Steadiness of the variational parameters}

To test the steadiness of the variational calculation, we have
studied the change of the variational parameters as a function of $x_{0}$, 
the vortex position. We have also checked the convergence of
these parameters in terms of the size of the numerical sample. Figure 
\ref{parameters} shows the behavior of the three variational parameters, the vortex
size $\zeta _{v}$, the Meissner parameter $\eta _{0}$, and
$\lambda _{M}$, as functions of the vortex position for different
sample sizes.

It can be seen in Fig. \ref{parameters}(a) that with an increase in the sample size, the last two parameters become independent of the vortex position, which is an indication of the adequacy of the variational function. On the other hand, the vortex parameter $\zeta _{v}$ is strongly affected
when the vortex moves close to the surface [Fig. \ref{parameters}(b)] and does not
show appreciable changes with increasing sample size. The value of $\zeta _{v}$ for large $x_o$ coincides with the value obtained by Clem for $\kappa=2$ in Ref. \onlinecite{clem}, $\zeta _{v}=1.15 \xi$.

The results of Fig. \ref{parameters} show that we can obtain the Meissner parameters
$\eta _{0}$ and $\lambda_{M}$ from the Meissner variational calculation and use
them as fixed parameters in the vortex variational equations. This
allows a faster convergence of the solution in the presence of
vortices because we only need to minimize the energy with respect to
a single parameter, the vortex core size $\zeta _{v}$. We have used
this procedure in what follows.

Our results show good agreement between the variational calculations and
full numerical results both for the energy barrier, as will be shown in
the next section, and for other quantities, particularly the shape
of the vortex near the surface.

\subsection{Quality of the variational solution of a vortex near a surface}

As we did in Sec. II for the Meissner state, in this section we
compare the variational solutions for a vortex near a surface to 
the results obtained from the full numerical solutions of the GL
equations. However, in the present case, the comparison needs to be more
carefully done. In the variational calculation, the position of the vortex at a
given point, $x_0$, is fixed and the energy must be minimized in
order to find the variational parameters. Due to the forces exerted
by the Meissner currents, such a configuration is unstable in the GL
case. The difficulties in pinning an isolated vortex at position $x_0$ are not solely related to the time dependent equations that we are using to find the equilibrium configurations. In a time independent approach, the system also tends to the equilibrium configuration that for a field larger than the field of first vortex penetration ($H_p$) is a vortex at position $x_0 \rightarrow \infty$.

In order to overcome this difficulty and to calculate the
energy of a vortex located at $x_0$, we pin the vortex by using a square
numerical seed of size $d=\xi$ for the order parameter. The use
of a pinning seed poses the problem of the distortion of the order
parameter around the vortex core, which affects the evaluation of the
free energy. We reduce this side effect by introducing a seed that 
has the same shape as a vortex in bulk. The solutions of the
numerical equations then converge towards a stable state containing
a vortex pinned at position $x_0$. We move the seed in steps given
by the spatial discretization, allowing the system to relax to a new
stable solution at each new position of the seed. Our choice of
pinning seed allows a good comparison to the variational
solution of a vortex at position $x_0$. A similar procedure was used in
Ref. \onlinecite{golib}, wherein by fixing the phase of the order parameter,
it was possible to pin and move vortices near a surface.

\begin{figure}
\begin{center}
\includegraphics[width=\linewidth]{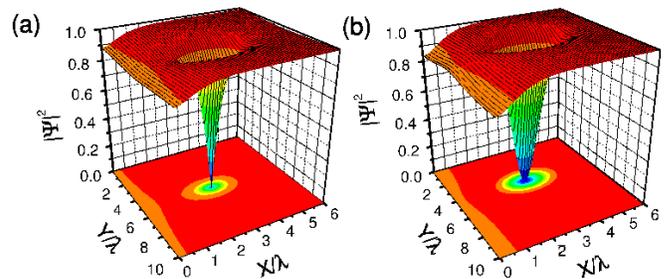}
\end{center}
\caption{(Color online) Comparison between the profiles of the order parameter from (a) the variational  model and (b) full numerical solution of the GL equations. Both graphs are for $\kappa=2$, $x_0=2.9\lambda$, and $H_a=0.70H_c$.} \label{profiles}
\end{figure}

\begin{figure}[htb]
\begin{center}
\includegraphics[width=\linewidth]{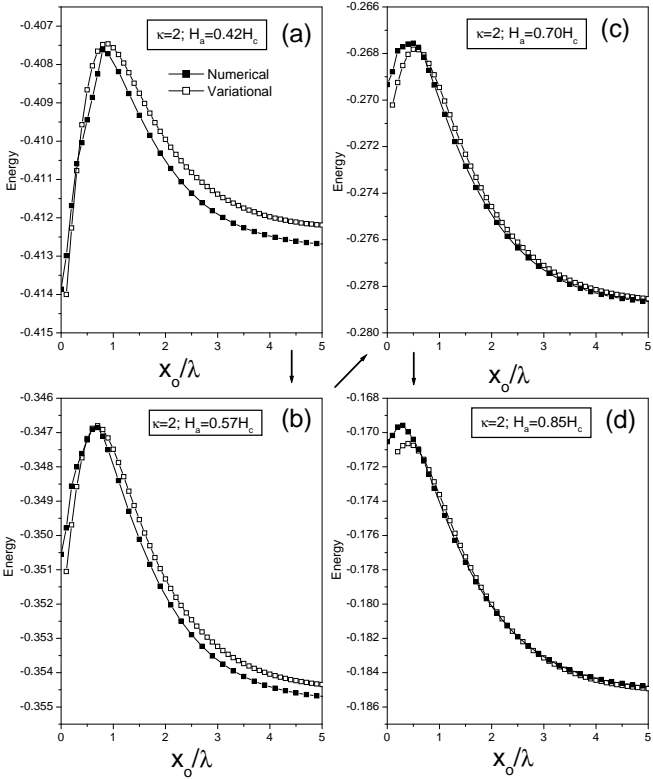}
\end{center}
\caption{Shown is a comparison between variational
calculation and numerical results for the energy barrier as a function
of the vortex position  for $\kappa =2$ and for different
values of the applied field. Arrows follow the increase in magnetic field from (a) to (d).
 It is seen that even  in (d), wherein $H_a$ is
near $H_{p}=1.04H_{c}$, the variational description is quite
accurate.} \label{kappa2}
\end{figure}

Figure \ref{profiles} shows a comparison between the profiles of the order parameter obtained from the variational method and from the full numerical simulations of the GL equations. The figures are for $x_0=2.9\lambda$ and $H_a=0.70H_c$. As we see from Figs. \ref{profiles}(a) and \ref{profiles}(b), the description obtained from the variational model is quite accurate, although some qualitative differences are apparent. This is a typical feature of variational calculations, wherein generally a better agreement is obtained for the energy calculation than for the field properties.

In Fig. \ref{kappa2}, we show a comparison between the variational calculations and numerical  results for the energy barrier as a function of the vortex position and for different values of the
applied magnetic field. The maxima of the energy as a function of the vortex position for fields
lower than $H_{p}$ generate the energy barrier for vortex penetration. The energy barrier can be defined as the energy difference between the value at the maximum and the value at the surface, $\Delta =G(x_{max})-G(0)$. By increasing the magnetic field from $H_a=0.42H_c$ [Fig. \ref{kappa2}(a)] to $H_a=0.70H_c$ [Fig. \ref{kappa2}(c)], it is seen that the energy
barrier decreases and, finally, almost disappears near $H_a=0.85H_c$, as shown in Fig. \ref{kappa2}(d). At the same time, the maximum moves closer to the surface.

It is seen that even when $H_a$ is near $H_{p}$, the variational description is quite
accurate [see Figs. \ref{kappa2}(c) and \ref{kappa2} (d)]. These results should be compared to those shown in Fig. 3 of Ref. \onlinecite{clem}, wherein it is seen that the energy of the vortex line quite accurately coincides with the full GL numerical results in a wide range of $\kappa$ values.

For $\kappa =2$, the field of first penetration obtained turns out to be $H_{p}=1.04H_{c}$, which is a value lower than the one obtained in Sec. II for the Meissner state ($H_{p}=1.13H_{c}$). In the full numerical simulations, this is a consequence of the symmetry breaking produced by the pinning center we have used; while in the variational approach, the symmetry is already broken by the nature of the solution we have imposed.

\begin{figure}[htb]
\begin{center}
\includegraphics[width=\linewidth]{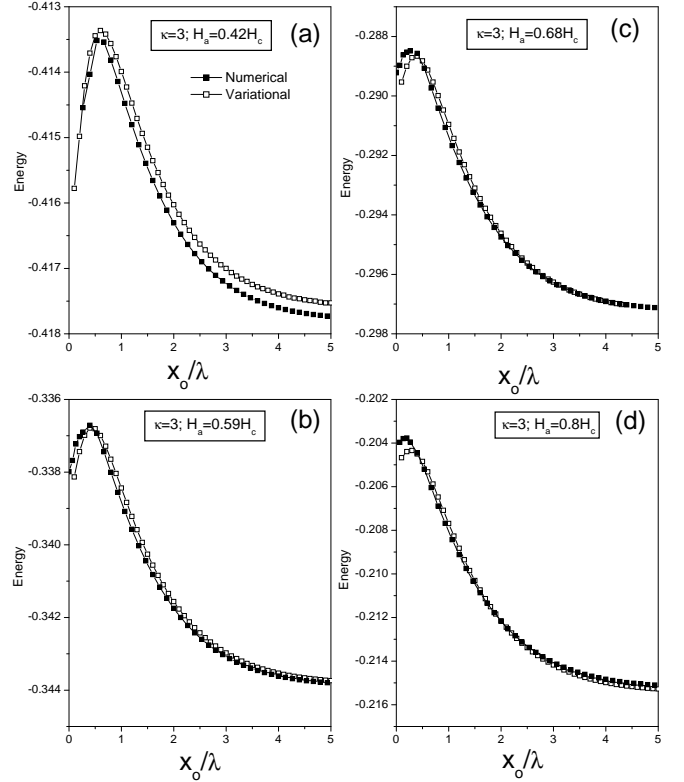}
\end{center}
\caption{Same as Fig. \ref{kappa2} but for $\kappa =3$. }
\label{kappa3}
\end{figure}

In Figs. \ref{kappa2}(c) and \ref{kappa2}(d), the energy differences between both approaches are higher near the surface, $x_0\approx 0$. This is a consequence of the increase in the force that the Meissner currents exert on the vortex. In the full numerical approach, this means a higher difficulty in pinning a vortex at position close to $x_0 \approx 0$. In any case, the energy differences are always lower than $0.05\%$.

Figure \ref{kappa3} shows a similar comparison to that in Fig. \ref{kappa2}
 in the case of $\kappa =3$. As can be seen from
Figs. \ref{kappa3}(c) and \ref{kappa3}(d), when $H_a$ is near
$H_{p}$, the variational description remains as accurate as in the previous case.
The maxima of the energy as a function of the vortex position for fields lower
than $H_{p}$ generates the energy barrier for vortex penetration.
By increasing the magnetic field from $H_a=0.42H_c$ [Fig. \ref{kappa3}(a)] to
$H_a=0.68H_c$ [Fig. \ref{kappa3}(c)], the energy barrier decreases and,
finally, almost disappears near $H_a=0.8H_c$, as shown in Fig. \ref{kappa3}(d).
For $\kappa =3$, the field of first penetration turns out to be $H_{p}=0.91H_{c}$.

\begin{figure}[htb]
\begin{center}
\includegraphics[width=\linewidth]{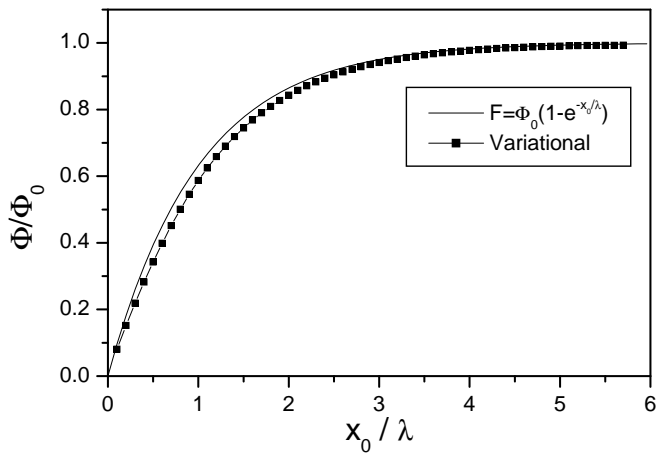}
\end{center}
\caption{Magnetic flux of a vortex as a function of the distance to the surface. The continuous curve corresponds to Eq. (\ref{flujo}) and the squares to the variational calculation.}
\label{flujoVsa}
\end{figure}

The usefulness of the variational approach can be stressed by
calculating other quantities related to vortices near surfaces. One
such quantity is the magnetic flux. As early as 1961, Bardeen
\cite{bardeen} showed that magnetic flux in a superconducting
cylinder can be less than one flux quantum. Later in Ref.
\onlinecite{shmidt}, Shmidt and Mkrtchyan, using an extension of the
London model, calculated the magnetic flux for a vortex near the
surface of a semi-infinite sample. They found the following
functional dependence:

\begin{equation}
\Phi=\int B \cdot da=\Phi_0(1-e^{-x_0}).
\label{flujo}
\end{equation}

We have used the variational model to calculate the magnetic flux as a function
of the distance to the sample surface, as shown in Fig. \ref{flujoVsa}.
Our results agree quite well with the functional dependence of Eq. (\ref{flujo}).
We note that the fluxoid quantization
\begin{equation}
\frac{2\pi}{\kappa}n=\Phi +\oint \frac{J_s}{|\Psi|^2}dl
\label{quantization}
\end{equation}
remains valid, even when $\Phi$ is less than one flux quantum, due to the
contribution of the superconducting currents $J_s$.

Geim et al., using a Hall probe in Ref. \onlinecite{geim1} experimentally confirmed 
the fact that vortices can have less than one flux
quantum in mesoscopic samples. Similar results had previously been obtained in 
experiments on bulk samples by Civale and de la Cruz
\cite{civale}.  In Ref. \onlinecite{civale}, they studied the
magnetic behavior as a function of temperature of samples with a
constant number of vortices pinned at fixed positions. They observed
that the magnetic flux carried by vortices located close to the
surface increases with decreasing temperature due to an indirect
increase in the distance to the surface when $\lambda(T)$ decreases.

\section{Extension to two vortices}

The variational model of Clem was previously extended to describe a flux lattice
in Ref. \onlinecite{haoclem}, wherein the reversible magnetization of high-T$_c$
superconductors as a function of the applied field was calculated.
Our formulation can be straightforwardly extended to the description of two or
more vortices near a surface. In particular, we focus on the case of two vortices,
located at a distance $x_0$ from the sample surface and separated by a distance
$y_0$. In order to calculate the interaction force, we first follow
the same procedure used in Sec. III to calculate the variational energy of the system.
We introduce an order parameter, which is the product of Clem's variational expressions
for the two vortices and the corresponding images. Similarly, the velocities and the total
magnetic field are obtained by following the procedure described in Sec. III.
Once the energy of the system containing two vortices is obtained, the interaction
force between them can be calculated from the numerical derivative of the energy of the system.

\begin{figure}[htb]
\begin{center}
\includegraphics[width=\linewidth]{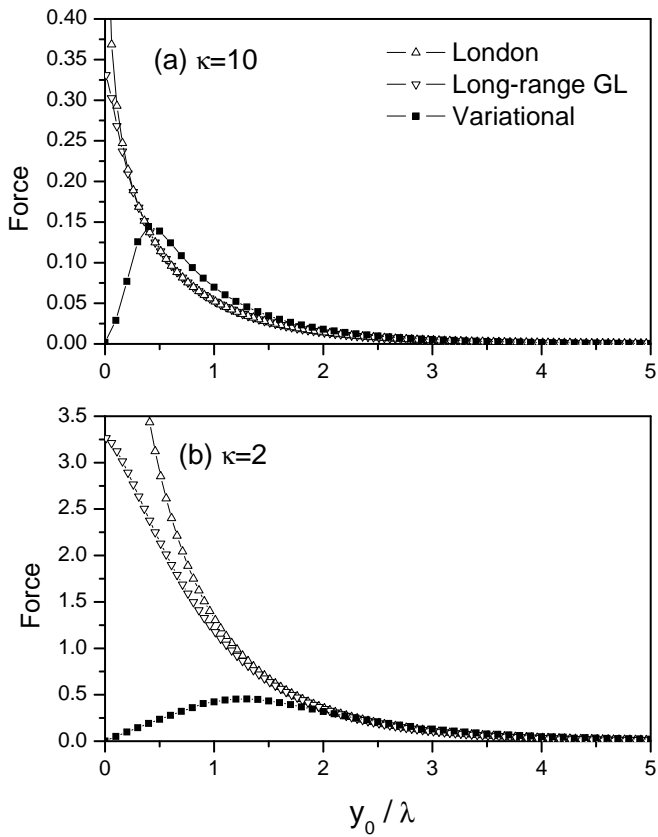}
\end{center}
\caption{Force between two vortices as a function of the distance between them when both vortices are well apart from the surface. In (a) and (b), we show a comparison of the behavior in the London ($\bigtriangleup$), long-range GL ($\bigtriangledown$), and variational ($ \blacksquare$) models.}
\label{flujoVsb}
\end{figure}

We concentrate first on the interaction force between vortices that are away from the
sample surface, i.e. for $x_0 \rightarrow \infty$. In this case, the image vortices
can be omitted from the calculations because their influence is negligible when they are far from the surface. In Fig. \ref{flujoVsb}, we show the interaction force for two cases:
Fig. \ref{flujoVsb}(a) is for $\kappa=10$ and Fig. \ref{flujoVsb}(b) is for $\kappa=2$.
A comparison is shown with the force calculated within the London model and also with the
formula obtained from the long-range asymptotic behavior  \cite{kramer, brandtreview}
within the Ginzburg--Landau approach. Good agreement between all curves is obtained for
distances larger than $0.5\lambda$ for $\kappa=10$ and for distances larger than $2.0\lambda$
for $\kappa=2$. It should be noted that in the case of the London model, the force corresponding
to two vortices $F_{L}=(4\pi /\kappa^2)K_0(r_{i}-r_j)$ diverges when the distance between
vortices $(r_{i}-r_j)$ tends to zero, which in our case is when $y_0 \rightarrow 0$.

The expression for the long-range asymptotic behavior of the GL
model, due to Kramer, is \cite{kramer, brandtreview}

\begin{equation}
F_{int}=\frac{4\pi}{\kappa^2}\{K_0(r_{i}-r_j)-K_0[\sqrt{2}\kappa(r_{i}-r_j)]\}.
\end{equation}

This expression incorporates a correction due to the overlapping of
the vortex cores that induces an attractive component in the force
between vortices. The final result gives a finite value for the
force corresponding to the long-range asymptotic GL model when $y_0
\rightarrow 0$, as can be seen in Figs. \ref{flujoVsb}(a) and
\ref{flujoVsb}(b).

On the other hand, the attractive contribution in the variational model results in a zero force for $y_0 = 0$ when the two vortices
merge in a two-quanta vortex. This configuration is unstable in a type II superconductor, and a minimal separation between vortices would result in a repulsive force. An interaction force decreasing and going to zero for very small
vortex separation is in agreement with previous variational calculations of the
interaction energy between vortices obtained by Jacobs and Rebbi in Ref. \onlinecite{jacobs}.
As we can see in Fig. \ref{flujoVsb}, the interaction force obtained from the
variational model has a maximum located at $y_0=0.4\lambda$ for $\kappa=10$ and at $y_0=1.3\lambda$ when $\kappa=2$.

\begin{figure}[htb]
\begin{center}
\includegraphics[width=\linewidth]{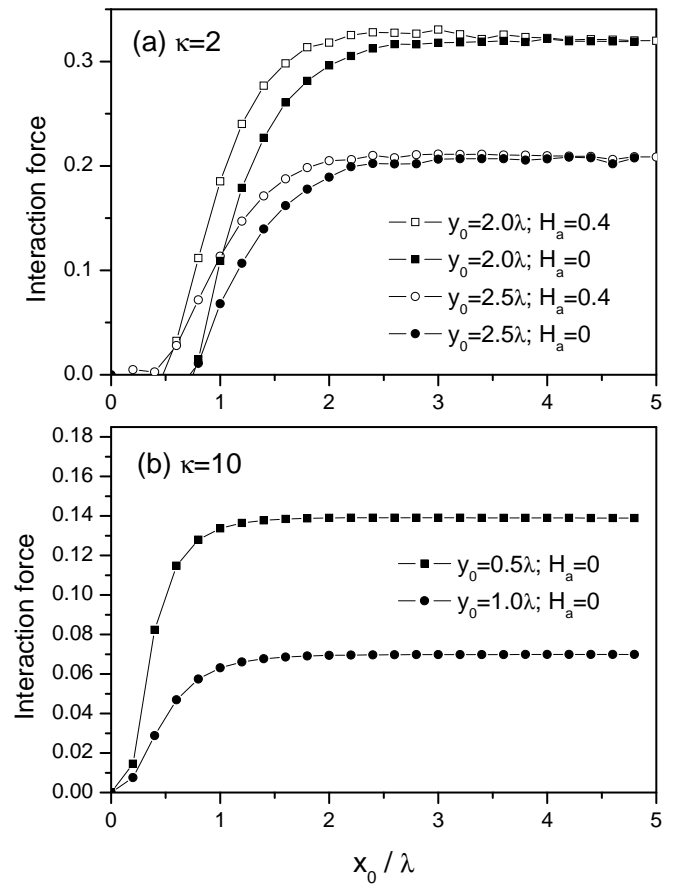}
\end{center}
\caption{Interaction force between two vortices separated a distance $y_0$ as a function of their distance to the surface $x_0$. In (a) and (b), we compare results obtained at different values of $\kappa$, $y_0$, and applied magnetic field $H_a$.}
\label{ForcVsx0}
\end{figure}

In Fig. \ref{ForcVsx0}, we fix the distance $y_0$ between two
vortices and calculate the variation of the interaction force
between them as a function of the distance $x_0$ to the surface. In
this case, the contribution of the image vortices becomes more
important as the pair approaches the surface. We show the
interaction force parallel to the surface (along the line that
connects both vortices) as a function of $x_0$ for $\kappa=2$ in
Fig. \ref{ForcVsx0}(a) and for $\kappa=10$ in Fig.
\ref{ForcVsx0}(b). In  Fig. \ref{ForcVsx0}(a), the open squares are
for $H_a=0.4$ and the closed squares are for $H_a=0$; in both cases, we
observe a steady decrease in the force between vortices when the
distance to the surface decreases. There is only a small difference
between the forces obtained at different values of the applied magnetic
field due to the differences between the Meissner currents induced in
each case. The overall qualitative behavior is independent of the
value of the distance between vortices $y_0$, as we see in Fig.
\ref{ForcVsx0}(a) by comparing the results for $y_0=2\lambda$ and
$y_0=2.5\lambda$. In Fig. \ref{ForcVsx0}(b), the same qualitative
behavior is observed for $\kappa=10$. From Fig. \ref{ForcVsx0}, we
can conclude that there is a steady decrease in the interaction
force between vortices when the distance to the surface decreases
and that the force goes to zero when the pair is close to the
boundary.

This result can be understood by considering that the repulsive interaction force between two vortices is screened by the contribution of the attractive interaction force between each vortex with the image of the other vortex. In particular, when $x_0 \rightarrow 0$, a vortex and its image are located at approximately the same position for the other vortex and the interaction force goes to zero.

Our results suggest that the vortex lattice is softer in the direction parallel to the surface in finite samples. It should be interesting to devise experiments that can explore these properties.
Even when we have calculated the case of a semi-infinite sample, the same qualitative behavior is expected to appear in a thin film with vortices parallel to the surface. In a thin film, vortices are confined by two surfaces and the image vortices corresponding to both surfaces contribute to the screening of the vortex interaction forces.

\section{Conclusions}

We have shown that Clem's variational ansatz for a free vortex can be extended
to the  description of vortex penetration. The results show quite good agreement
with the full numerical results both for the energy barrier and for the description
of the vortex near the surface. The flux carried by a vortex as a function of its
distance to the surface can be shown to be easily calculated and to coincide
with known results.

We extended the model to calculate the force between two vortices.
When the vortices are far from the surface, the variational
results show good agreement with the London and long-range GL
results for large intervortex distances; whereas for small distances,
the variational model gives vanishing forces corresponding to the
merging of the two vortices in a double quantized vortex \cite{escoffier}. We also found a steady decrease in the interaction force between vortices when the distance to the surface decreases; the interaction force goes to zero when the pair is close to the boundary.

Our variational approach gives manageable expressions that can be used to obtain
approximations for all physically relevant quantities. Another advantage of this
method is the lower computational time that it requires, allowing one to obtain fast
and reliable solutions for a vortex near a surface. The agreement between the
variational solution and numerical calculations shows the usefulness of the former
for intermediate  $\kappa$ when computations become heavy. For large $\kappa$, numerical
calculations can be based on the London model for the magnetic contribution.
This description is not accurate at lower $\kappa$ values where a numerical
approach to the GL description is more appropriate. A particular problem arises at intermediate $\kappa$, at which the computation within the GL model becomes very
demanding because of the difficulty to describe both spatial scales with the
same discretization. It is in this range that the variational
approach is most welcome.

The method used in this paper can be generalized for mesoscopic
superconductors. However, consideration of more than one surface in 
some cases could give rise to infinite images. This difficulty can be overcome by 
truncating the infinite series as was done in Ref. \onlinecite{mingo2} for
vortex penetration in a thin film using the London model.
In this paper, we have assumed a semi-infinite medium with no demagnetizing effects. In this case, the boundary condition for the magnetic field ${\bf B}|_{s}={\bf H}_{a}$ applies at the sample surface. A similar condition applies for a thin film with the externally applied magnetic field parallel to the surface. A thin film is an example of a mesoscopic system wherein our results can be generalized in a straightforward manner. However, in general, demagnetization effects are important in mesoscopic superconductors of finite thickness where the boundary condition ${\bf B}|_{s}={\bf H}_{a}$ applies at infinity and not at the sample boundary. Vortices in mesoscopic superconductors are confined by the sample surface and their interaction with the Meissner currents is very important, a situation with similarities to the case analyzed in this work. This opens up interesting questions about the behavior of the effective interaction force between two vortex cores as a function of the distance to the surface in mesoscopic superconductors.

\section{Acknowledgments}

We are grateful to Daniel Dom\'{\i}nguez and F. de la Cruz for their valuable
suggestions and Jorge Berger for carefully reading the manuscript. 
We acknowledge support from CNEA, CONICET (Grant No. PIP 5596),
and ANPCyT (Grants No. PICT 13829 and No. PICT 13511).

\end{document}